\documentclass{iau} 
\usepackage{graphicx}

\title[X-ray detected far-IR AGN in the green valley] %% give here short title %%
{Properties of X-ray detected far-IR AGN in the green valley}

\author[ Mahoro et al.]   %% give here short author list %%
{Antoine Mahoro $^{1,\,2}$, Mirjana Povi\'c$^{3,\,4}$, Petri V\"ais\"anen$^{1,\,5}$, Pheneas Nkundabakura$^{6}$, Beatrice Nyiransengiyumva$^{6}$, and Kurt van der Heyden$^{2}$}

\affiliation{$^{1}$South African Astronomical Observatory, P.O. Box 9 Observatory, Cape Town, South Africa\\
$^{2}$Department of Astronomy, University of Cape Town, Private Bag X3, Rondebosch 7701, South Africa\\
$^{3}$Ethiopian Space Science and Technology Institute (ESSTI), Entoto Observatory and Research Center (EORC), \\
Astronomy and Astrophysics Research Division, P.O. Box 33679, Addis Ababa, Ethiopia\\
$^{4}$Instituto de Astrof\'isica de Andaluc\'ia (IAA-CSIC), Glorieta de la Astronom\'ia s/n, 18008 Granada, Spain\\
$^{5}$Southern African Large Telescope, P.O. Box 9 Observatory, Cape Town, South Africa\\
$^{6}$MSPE Department, School of Education, College of Education, University of Rwanda,  P.O. Box 5039, Kigali, Rwanda}

\pubyear{2019}
\volume{356}  %% insert here IAU Symposium No.
\jname{Nuclear Activity in Galaxies Across Cosmic Time}
\editors{M. Povi\'c, P. Marziani, J. Masegosa, H. Netzer,\\ S. H. Negu, \&
	S. B. Tessema, eds.}
	
\begin{document}
\maketitle

\begin{abstract}
In this study, we analysed active galactic nuclei in the “green valley” by comparing  active and non-active galaxies  using data from the COSMOS field.
We found that most of our X-ray detected active galactic nuclei with far-infrared emission have star formation rates higher than the ones of normal galaxies of the same stellar mass range.

\end{abstract}

\firstsection % if your document starts with a section,
\section{Introduction}

In optical colour-magnitude diagrams (CMDs) galaxies predominantly lie along either the “red sequence,” characterised with red colours and elliptical or spheroidal morphologies, or in the “blue cloud,” characterised by blue colours and disk or irregular morphologies.
The differences between these two groups create a bimodality in the color-magnitude distribution of galaxies. The two populations have a nontrivial region of overlap between them termed the "green valley" (Wyder et al. 2007; Salim et al. 2009), through which past attempts have been made to place a quantitative divider that splits the two populations on the basis of colour.

Previous works that studied green valley galaxies suggested that they are a transitional phase between the blue cloud and red sequence in terms of star formation, colours, stellar mass, luminosity, and different morphological parameters (Povi\'c et al. 2012; Schawinski
et al. 2014; Salim 2014; Lee et al. 2015; Phillipps et al. 2019).

Interestingly, the rate of active galactic nuclei (AGN) detection is high in green valley galaxies, whether AGN are selected by deep X-ray surveys (Nandra et al. 2007; Coil et al. 2009; Povi\'c et al. 2012) or by optical line-ratio diagnostics (Salim et al. 2007).

Previous works proposed that AGN negative feedback plays a key role in the galaxies’ process of quenching star formation (SF) and galaxy transformation. In this work, we went one step further by  looking at the transition processes from the blue cloud to the red sequence on one side, and at AGN triggering mechanisms and their connection with normal galaxies on the other, by using AGN and normal galaxies in the green valley.

\section{Data}
In this work, we used the data from COSMOS survey, aimed at studying the formation and evolution of galaxies as a function of both, cosmic time and the local galaxy environment (Scoville et al. 2007).

Our main catalogue was taken from the Tasca et al. (2009). In order to select AGN we used the ratio between the X-ray flux ($\rm{F_{x}}$) in the hard 2–10 keV band and the optical \textit{i} band flux ($\rm{F_{o}}$) (Bundy et al. 2007; Trump et al. 2009). Non-AGN galaxies were selected by removing sources selected as AGN, which is explained in Mahoro et al. (2017).

There have been many ways of defining the green valley galaxies, all being based on the bimodal distribution of galaxies when using different colours. In this work, we used the U - B rest-frame colour and  criteria $\rm{0.8\leq U-B\leq 1.2}$ (e.g. Nandra et al. 2007). To obtain U-B rest-frame colours, we first run KCORRECT code (Blanton \& Roweis 2007) to apply the k-correction on both CFHT \textit{u} and Subaru B bands.
The far-infrared data was obtained by  cross-matching  the previously obtained sample with Herschel/PACS data as explained in Mahoro et al. (2017).

\section{Star formation rates (SFR)}

We determine the SFRs of the FIR selected green valley AGN and non-AGN galaxies using the spectral energy distribution (SED)-fitting method the LEPHARE code (Ilbert
et al. 2006) and by assuming that all the FIR luminosity is due to star formation for non-activate galaxies. To measure SFRs we use Kennicutt et al. (1998) relation, as explained in Mahoro et al. (2017). To fit AGN we use Kirkpatrick et al. (2015) templates with known AGN contribution to the IR luminosity.  Non-AGN galaxies were fitted using Chary \&
Elbaz (2001) libraries.

\begin{figure}[htp]
\centering
\includegraphics[width=8cm,height=8cm]{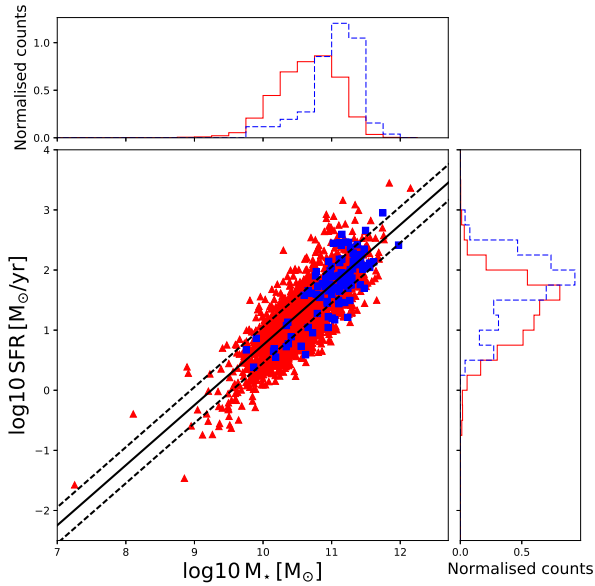}
\caption{SFR vs. stellar mass. AGN are represented by blue squares, while non-AGN are represented by red triangles. The solid black line shows the Elbaz et al. (2011) fit for the MS, while the dashed lines represent the MS width of $\pm$0.3 dex. Top and right histograms show the normalised distributions of stellar mass and SFR respectively, and comparison between the AGN (blue dash lines) and non-AGN (red solid lines) samples.
 \label{MS_both}}
\end{figure}

 In Fig. \ref{MS_both} we present the relation between SFR and stellar mass (e.g. Netzer et al. 2016;
Povi\'c et al. 2016), and study the location of our AGN and non-AGN sources in relation to the main sequence of  star-forming galaxies taken from Elbaz et al. (2011).
  
Our sample of FIR green valley AGN do not show signs of SF quenching (as suggested in the majority of previous studies of X-ray detected AGN) since 68\% and 14\% of of sources are located on and above the MS of SF, respectively. On the other hand we have 70\% of non-AGN FIR green valley galaxies on the MS, with 9\% being above it. 
\section{Morphological classification and analysis}
\begin{figure}[htp]
\centering
 \includegraphics[width=8cm,height=8cm]{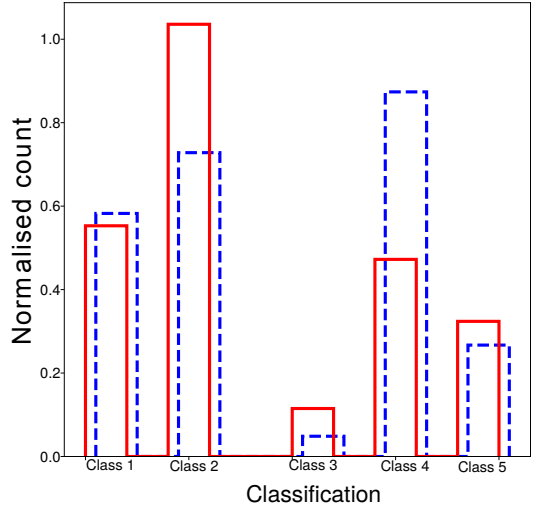}
 \caption{Normalised distributions of visually classified morphological types of FIR green valley AGN (blue dashed lines) and non-AGN (red solid lines).}
 \label{Visual_class}
\end{figure}

We also carried out the morphological analysis of selected FIR AGN and FIR non-AGN green valley galaxies to understand better their location with respect to the main sequence of star formation obtained in Mahoro et al. (2017). By using the HST/ACS F814W images, we did visual morphological classification of all FIR AGN and non-AGN galaxies samples. Classification was done by three independent classifiers, separating all galaxies into:
\begin{itemize}
\item class 1: elliptical, S0 or S0/S0a,
\item class 2: spiral,
\item class 3: irregular,
\item class 4: peculiar and 
\item class 5: unclassified.
\end{itemize}
Fig. \ref{Visual_class}  shows the normalised distribution for final visual classification. We can clearly see a difference in class 4, with 38\% for FIR AGN and 19\% of
non-AGN galaxies being peculiar, with clear signs of interactions and mergers. On the other hand, we also obtained an important difference in class 2, finding higher fractions of FIR non-AGN (46\%) in comparison to AGN (26\%). We compared our visual classifications with available non-parametric classifications in COSMOS, and we did more analysis of the distributions of different morphological parameters by comparing FIR AGN and FIR non-AGN green valley galaxies.  We found the standard behaviour of morphological parameters, as explained in Mahoro et al. (2019).

\section{Discussion}
\begin{figure}[htp]
\centering
\begin{minipage}[c]{0.49\textwidth}  
 \includegraphics[width=7.5cm,angle=0]{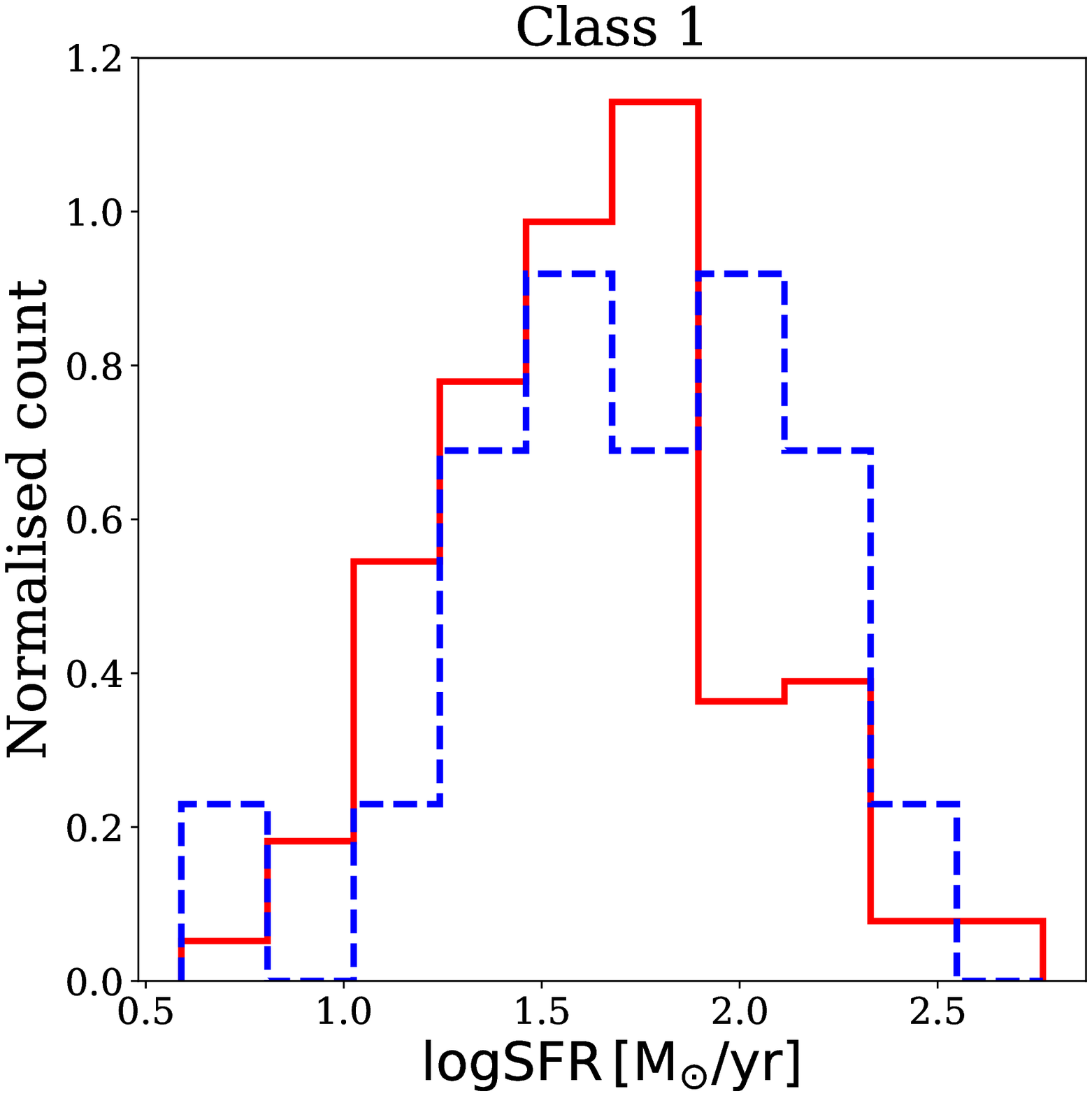}
\end{minipage}
\begin{minipage}[c]{0.49\textwidth}
\includegraphics[width=7.5cm,angle=0]{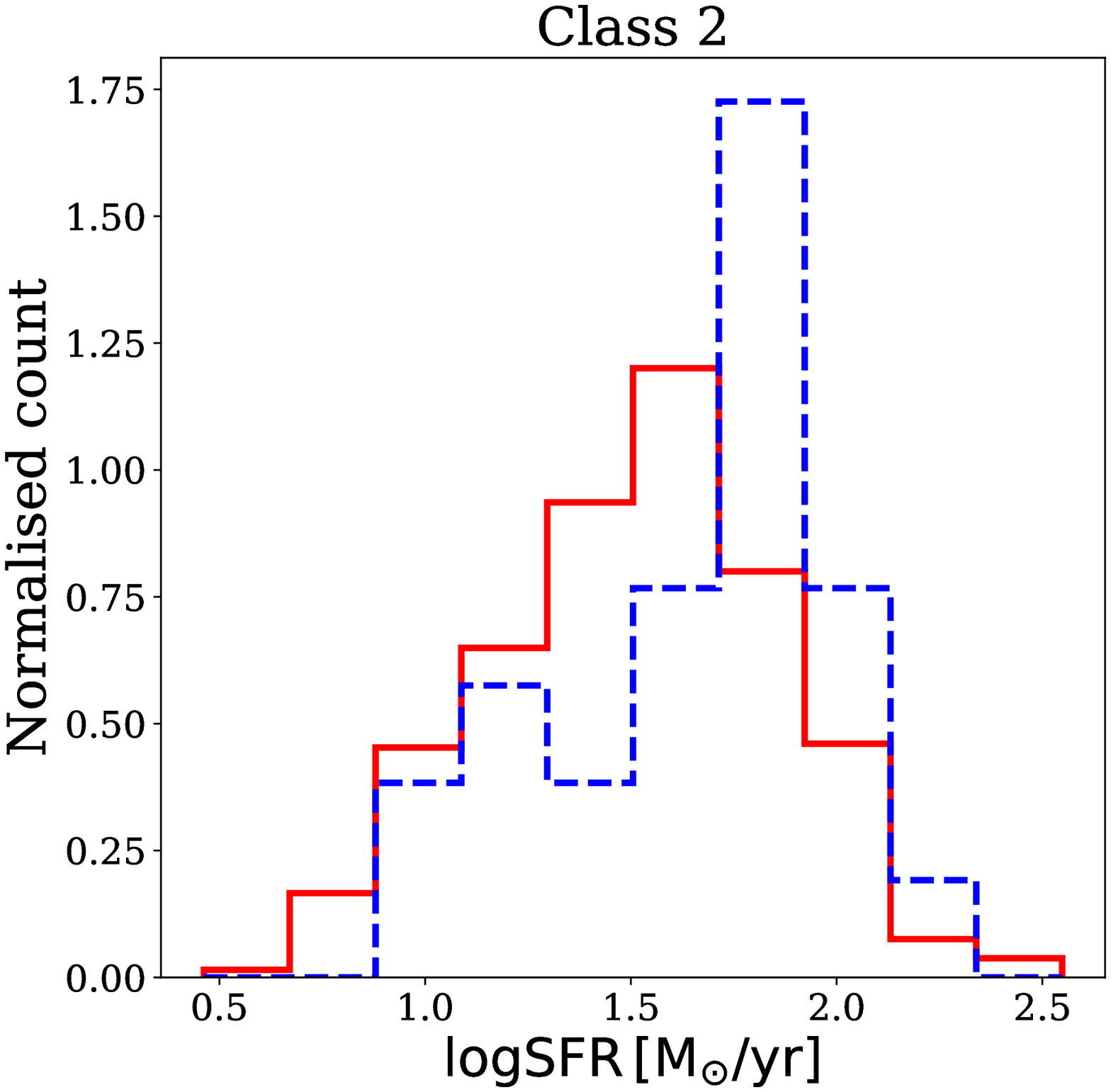}
\end{minipage}
\begin{minipage}[c]{0.49\textwidth}
\includegraphics[width=7.5cm,angle=0]{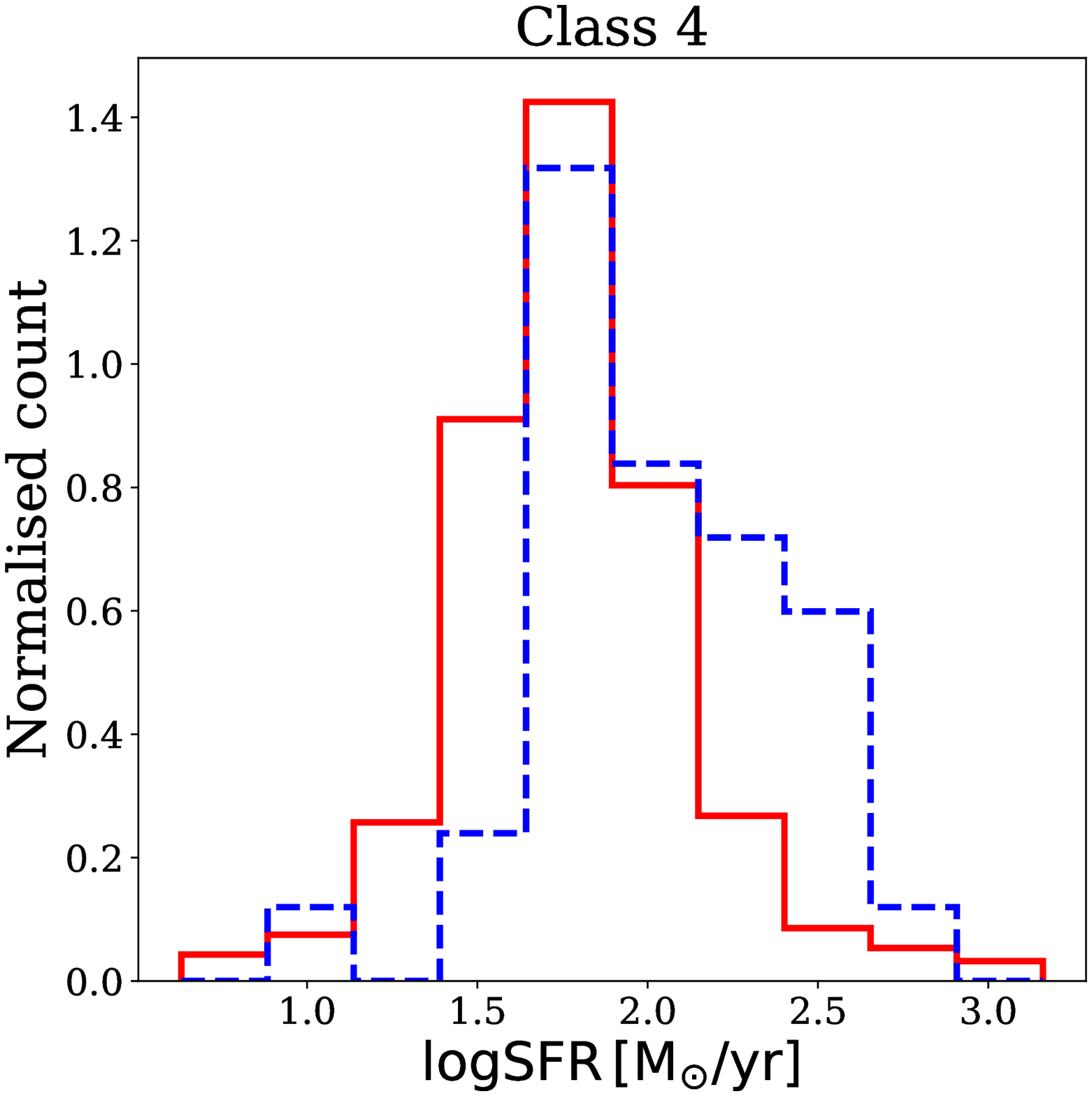}
\end{minipage}
\caption{Normalised distributions of SFR of FIR AGN (blue dashed histograms) and non-AGN (red solid histograms) for a fixed stellar mass range of logM*\,=\,10.6M$_\odot$\,-\,11.6M$_\odot$ in relation to morphology.}
\label{fig_SFR_SameMAssRange}
\end{figure}

We found that FIR AGN have higher SFRs in comparison to FIR non-AGN green valley galaxies. Fig. \ref{fig_SFR_SameMAssRange} shows the comparison of SFRs of AGN and non-AGN samples for the same stellar mass ranges taking into account morphology.  In all cases the AGN sample shows higher SFRs in comparison to non-AGN galaxies, independently of morphology (Mahoro
et al. 2019). In case of class 4, interactions and mergers could explain higher values of SFRs observed in AGN. However, taking into account higher SFRs in the rest of AGN green valley sample (e.g., in 26\% and 25\% of class 1 and class 2 galaxies, respectively), we suggested that interactions and mergers alone cannot explain the results of Mahoro et al. (2019). Therefore, if there is an impact of AGN feedback on star formation in case of selected FIR emitters it looks to be rather positive than negative. 
\section*{Acknowledgments}
This work was supported by the National Research Foundation of South Africa (Grant Numbers 110816). AM acknowledges financial support from the Swedish International Development Cooperation Agency (SIDA) through the International Science Programme (ISP) - Uppsala University to University of Rwanda through the Rwanda Astrophysics, Space and Climate Science Research Group (RASCSRG), East African Astro- physics Research Network (EAARN) are gratefully acknowledged. MP acknowledges financial supports from the Ethiopian Space Science and Technology Institute (ESSTI) under the
Ethiopian Ministry of Innovation and Technology (MoIT), and from the Spanish Ministry of Science, Innovation and Universities (MICIU) through projects AYA2013-42227-P and AYA2016-76682C3-1-P. PV acknowledges support from the National Research Foundation of South Africa.

%\bibliographystyle{aa}
%\bibliography{Ref_proc} 
\end{document}